\documentclass[A4,10pt]{article} 

\usepackage{opticameet3} 
\usepackage[longtable]{multirow}
\usepackage[utf8]{inputenc} 
\usepackage[T1]{fontenc}    
\usepackage{hhline}
\usepackage{graphicx}       
\usepackage{float}
\usepackage{longtable}
\usepackage{epstopdf}
\usepackage[table,xcdraw]{xcolor}
\usepackage{tabularray}
\usepackage[pdf]{pstricks}
\usepackage{array}
\usepackage{rotating}
\usepackage{lscape}
\usepackage{colortbl} 
\usepackage[width=\textwidth]{caption}
\newcommand\authormark[1]{\textsuperscript{#1}}

\usepackage{amsmath,amssymb}
\usepackage[colorlinks=true,bookmarks=false,citecolor=blue,urlcolor=blue]{hyperref} 

\begin{document}

\title{Towards a deep learning approach for classifying treatment response in glioblastomas}


\author{Ana Matoso\authormark{1,2,*}, Catarina Passarinho\authormark{1}, Marta P. Loureiro\authormark{1}, José Maria Moreira\authormark{2}, Pedro Vilela\authormark{3}, Patrícia Figueiredo\authormark{1}, Rita G. Nunes\authormark{1}}

\address{\authormark{1}Institute for Systems and Robotics - Lisbon and Department of Bioengineering of Instituto Superior Técnico, Lisbon, Portugal\\
\authormark{2}Hospital da Luz Learning Health, Lisbon, Portugal\\
\authormark{3} Imaging Department, Hospital da Luz, Lisbon, Portugal\\}

\email{\authormark{*}anamatoso@tecnico.ulisboa.pt} 


\begin{abstract}
Glioblastomas are the most aggressive type of glioma, having a 5-year survival rate of 6.9\%. Treatment typically involves surgery, followed by radiotherapy and chemotherapy, and frequent magnetic resonance imaging scans to monitor disease progression. To assess treatment response, radiologists use the Response Assessment in Neuro-Oncology (RANO) criteria to categorize the tumor into one of four labels: complete response, partial response, stable disease, and progressive disease. This assessment is very complex and time-consuming. Our work aimed to implement the first Deep Learning pipeline for the classification of RANO criteria.
The models were trained and tested on the open dataset LUMIERE which has 638 timepoints of 91 patients. Five-fold cross-validation was performed using an 80/20 stratified split. Five approaches were sequentially evaluated: 1) inputting subtraction images instead of individual timepoints, 2) using different combinations of imaging modalities, 3) implementing different model architectures, 4) employing different pretraining tasks, and 5) adding patient clinical data to the inputs. The best-performing pipeline used a Densenet264, considering only T1-weighted, T2-weighted, and Fluid Attenuated Inversion Recovery images as input without any pretraining. A median Balanced Accuracy of 50.96\% was achieved (over 55\% in two folds). Additionally, two explainability methods were applied: Saliency Maps, and Grad-CAM (Gradient-Weighted Class Activation Mapping). These results set a benchmark for studies on glioblastoma treatment response assessment using RANO criteria while emphasizing the heterogeneity of factors that play a role when assessing glioblastomas’ response to treatment.

\end{abstract}

\section{Introduction}
Gliomas are a type of brain tumor that originates in the glial cells of the central nervous system and comprise 80\% of all malignant brain tumors \cite{schwartzbaum_epidemiology_2006}. The World Health Organization (WHO) organizes gliomas into four grades, being high-grade gliomas the most aggressive ones. Glioblastomas, characterized by the IDH (Isocitrate Dehydrogenase) wild-type genetic marker, are grade 4 high-grade gliomas and, as such, have a very poor prognosis, having a 1-year survival rate of just 42.7\% and a 5-year survival rate of 6.9\% \cite{ostrom_cbtrus_2022,Torp2022Sep}. These tumors are usually diagnosed using magnetic resonance (MR) images, and their treatment consists of surgery for maximum safe resection of the tumor followed by radiotherapy and chemotherapy (with temozolomide) \cite{stupp_radiotherapy_2005}. These patients then usually undergo frequent magnetic resonance imaging (MRI) scans to track disease progression \cite{horbinski_nccn_2023}. Nevertheless, due to the infiltrative nature of glioblastomas, around 90\% of patients develop recurrent tumors in the brain within 2 years of the initial diagnosis. This occurs sometime after surgery and may happen in the same region as the original tumor or in another more distant region \cite{mohammed_survival_2022, hoggarth_clinical_2024}.

Hence, measuring the response of these tumors to treatment becomes essential to deciding whether to maintain the same treatment strategy or to change approaches. The Response Assessment in Neuro-oncology (RANO) criteria \cite{wen_updated_2010} divide the treatment response into four categories based on imaging and clinical features: complete response (CR), partial response (PR), stable disease (SD), and progressive disease (PD). To classify the tumor into one of these classes, the neuroradiologist needs to assess the volume changes of the lesion throughout time, sometimes based on more than one MRI modality. Critically, 20-30\% of patients who undergo radiotherapy experience a phenomenon called pseudoprogression at the beginning of treatment. This refers to visual signs of apparent progression in the first MRI scan post-treatment, which disappear in later scans without the need to change the treatment plan \cite{brandsma_clinical_2008,taal_incidence_2008}. Pseudoprogression complicates the physician's assessment of the patient’s response to treatment. Therefore, there is a need for a better and faster assessment of tumor response to treatment. 

Artificial intelligence models have been used to tackle many medical imaging problems \cite{singh_how_2022, zlochower_deep_2020, shaver_optimizing_2019, lundervold_overview_2019} either with classical machine learning (ML), which uses imaging features to learn the outcomes (as in radiomics), or resorting to deep learning (DL), which intrinsically performs feature engineering, avoiding the need for explicit feature computation and selection. The use of radiomics comes with the added challenge of defining the regions of interest and segmenting them, which is in itself a process that researchers have been trying to automate \cite{singh_3d_2020, ghaffari_automated_2020}. Thus, in this work, we bypass the delineation of regions of interest and investigate DL approaches for the classification of glioblastoma treatment response into one of the four RANO criteria using MR images and clinical data. We also aim to investigate whether this approach could help identify regions outside the tumor suffering from infiltration, not yet visible to the naked eye. To the authors’ knowledge, this is the first paper that reports an automated classification of RANO criteria.

\subsection{State of the Art}

With the development of software and hardware tools, the research community has created several tools for neuroimaging analysis. However, only in the last decade has there been a leap in the development of artificial intelligence-based tools. This can be attributed to the advancement in computing power allied to the community challenges that have been created, such as the Brain Tumor Segmentation (BraTS) challenge \cite{ghaffari_automated_2020} (which was first run in 2012).

One important bottleneck when training DL models is the amount of data needed for the model to effectively learn and, even more importantly, its quality. The BraTS challenge is perhaps the most well-known competition, as it released a considerable amount of multiparametric MR images accompanied by ground-truth segmentation labels that were delineated by expert neuroradiologists. These kinds of datasets are of great importance since they cover two big pillars for the development of AI models: many data samples (which increase performance) and high-quality labels (which increase the model's reliability). Recently, two BraTS challenges (2020 and 2021) \cite{noauthor_brain_2020, baid_rsna-asnr-miccai-brats-2021_2023} have also included survival data for a regression task and the methylation status of the O6 methylguanine-DNA methyltransferase (MGMT) gene promoter for a classification task. 

Saxena et. al. \cite{saxena_fused_2023} studied different frameworks for classifying MGMT methylation status using the BraTs 2021 dataset, including classical machine learning techniques, DL, and a fusion of both (where the features from the ML models are integrated in the DL architecture). With the fusion framework, they were able to improve the performance of the classifier (accuracy) from 61\% using just ML and 69\% using just DL to 76\%.

Moreover, private datasets have also been used for classification tasks. Although the use of such datasets hinders reproducibility, some of these studies have achieved interesting results. Moassefi and Faghani et al. \cite{moassefi_deep_2022} trained a DL model based on a densenet network architecture to distinguish between pseudoprogression and true progression, reporting a mean accuracy of 76.4\% in a cohort of 124 patients.

Regarding the evaluation of tumor response, there have been several studies developing models to distinguish true progression from pseudoprogression/treatment related changes. Jang et al \cite{jang_prediction_2018} and Bacchi et al \cite{Bacchi2019Dec}, though with a limited amount of closed-source data, have obtained promising Area Under the Receiver Operating Characteristic Curve (AUC) values of 0.83 and 0.82, respectively. The first study employed a CNN-LSTM (Convolutional Neural Network with a Long Short Term Memory block) model and paired MRI data (a predetermined number of slices, of T1 and contrast-enhanced T1) and clinical data from 78 patients, whilst the later study achieved the best results using a combination of DWI (Diffusion Weighted Imaging) and Fluid Attenuated Inversion Recovery (FLAIR) images as input to a 3D CNN model. Li et al. \cite{Li2020Mar} and Ortega-Martorell et al. were able to achieve very high AUC/accuracy values (0.947/0.92 and 0.99/0.99, respectively): the first one by introducing a novel feature learning method based on deep convolutional generative adversarial network and AlexNet; and the second by using MRSI (Magnetic Resonance Spectroscopic Imaging) and 1D-CNN in mice. More recently, Gomaa et al \cite{Gomaa2025May} were able to reach an AUC of 0.75 in an external dataset using a visual transformer and contrast-enhanced T1 (CT1) and FLAIR images coupled with the corresponding clinical data.

Some researchers have investigated the segmentation of the tumor region followed by the calculation of the volumetric and bidimensional measurements according to the RANO guidelines using the BraTS dataset (with post-hoc annotations from hired experts) and private datasets. Namely, Chang et al. \cite{chang_automatic_2019} developed a fully automated algorithm that segments FLAIR and contrast-enhanced T1-weighted images for the estimation of hyperintensity volumes as well as the product of the maximum bidimensional diameters; their algorithm appeared to be fairly robust with an intraclass correlation coefficient over 0.90. Likewise, Nalepa et al. \cite{nalepa_deep_2023} expanded upon this work, detailing an end-to-end pipeline for the calculation of bidimensional RANO measurements and correlated these measurement with the tumor volume acquired with an automatic algorithm showing a strong correlation. In Table \ref{tab:lit-review} we present a summary of these studies, namely their goals, methods and results.

\begin{sidewaystable}[htbp]
\centering
\caption{Summary of prior work on progression classification in gliomas. OS=Open-Source, CS=Closed-Source, CT1=contrast-enhanced T1, FLAIR=Fluid Attenuated Inversion Recovery, AUC=Area Under the Receiver Operating Characteristic Curve, MRI=Magnetic Resonance Imaging, MRSI=Magnetic Resonance Spectroscopic Imaging, ML=Machine Learning, CNN=Convolutional Neural Network, Grad-CAM=Gradient-Weighted Class Activation Mapping, ICC=Intraclass Correlation Coefficient, RANO=Response Assessment in Neuro-oncology, DTI=Diffusion Tensor Imaging, AutoRANO=Product of maximum bidimensional diameters according to the RANO criteria, DWI=Diffusion Weighted Imaging, ADC=Apparent Diffusion Coefficient, LSTM=Long Short Term Memory}
\label{tab:lit-review}
\begin{tabular}{|>{\hspace{0pt}}m{0.181\linewidth}|>{\hspace{0pt}}m{0.062\linewidth}|>{\hspace{0pt}}m{0.165\linewidth}|>{\hspace{0pt}}m{0.267\linewidth}|>{\hspace{0pt}}m{0.262\linewidth}|} 
\hline
Title                                                                                                                                     & Reference                    & Goal                                                                                                                             & Methods                                                                                                                                                                                                         & Results                                                                                                                                                                                                   \\ 
\hline
A self-supervised multimodal deep learning approach to differentiate post-radiotherapy progression from pseudoprogression in glioblastoma & Gomaa et al, 2025            & Distinguish pseudoprogression from true progression                                                                              & 180 MRIs (OS, CT1 and FLAIR), corresponding clinical data and a pretrained Visual transformer were used as well as an external dataset (n=20)                                                                   & AUC=0.753 on external dataset with accuracy=0.750 using multimodal data                                                                                                                                   \\ 
\hline
Tracking Therapy Response in Glioblastoma Using 1D Convolutional Neural Networks                                                          & Ortega-Martorell et al, 2023 & Classify normal brain vs tumurous and unresponsive tumor vs responsive tumor                                                     & 2525 voxels from mice MRSI (CS); classical ML and 1D-CNNs were tested; used Grad-CAM                                                                                                                            & best model (1D-CNN) achieved over 99\% accuracy, specificity and F1-score                                                                                                                                 \\ 
\hline
Deep learning automates bidimensional and volumetric tumor burden measurement from MRI in pre- and post-operative glioblastoma patients   & Nalepa et al, 2023           & Segment tumor regions and calculate volumetric and bidimensional measurement according to RANO guidelines                        & 760 pre-operative and 504 post-operative MRIs (OS and CS, CT1, T1, T2, FLAIR); Segmentation performed using an ensemble of confidence-aware nnU-Nets; RANO measurements calculated using a rule-based algorithm & Automatic segmentation method had ICC0.7 in volumetric measurements when compared to experts. Bidimentional measurements stongly correlate with enhancing tumor region volume                             \\ 
\hline
A deep learning model for discriminating true progression from pseudoprogression in glioblastoma patients                                 & Moassefi et al, 2022         & Distinguish pseudoprogression from true progression                                                                              & 124 MRIs (CS, CT1 T1. T2, FLAIR) and a pretrained Densenet121 were used                                                                                                                                         & Accuracy=0.764 and AUC=0.756                                                                                                                                                                              \\ 
\hline
DC-AL GAN: Pseudoprogression and true tumor progression of glioblastoma multiform image classification based on DCGAN and AlexNet         & Li et al, 2020               & Distinguish pseudoprogression from true progression                                                                              & 84 DTI MRIs (CS) and deep convolutional generative adversarial network was used                                                                                                                                 & AUC=0.947 with accuracy=0.92                                                                                                                                                                              \\ 
\hline
Automatic assessment of glioma burden: a deep learning algorithm for fully automated volumetric and bidimensional measurement             & Chang et al, 2019            & Assess repeatibility of automatic hyperintensity volume calculation; Compare manual and automatic segmentations of tumor volumes & 843 pre-operative MRIs and 713 post-operative MRIs (OS and CS); Used a 3D U-Net for segmentation; RANO measurements calculated using a rule-based algorithm                                                     & FLAIR hyperintensity volume, contrast-enhancing volume, and AutoRANO were highly repeatable for the double-baseline visits, with ICC0.97 and manual and automatically measured tumor volumes had ICC0.84  \\ 
\hline
Deep learning in the detection of high-grade glioma recurrence using multiple MRI sequences: A pilot study                                & Bacchi et all, 2019          & Distinguish recurrence from treatment related changes                                                                            & 55 MRIs (CS, CT1, FLAIR, DWI and ADC maps) and a 3D CNN were used                                                                                                                                               & Best performance using DWI+FLAIR with AUC=0.8 and accuracy=0.82                                                                                                                                           \\ 
\hline
Prediction of Pseudoprogression  versus Progression using Machine  Learning Algorithm in Glioblastoma                                     & Jang et al, 2018             & Distinguish pseudoprogression from true progression                                                                              & 78 MRIs (CS, CT1, T1), corresponding clinical data and a CNN-LSTM were used                                                                                                                                     & Best model achived AUC=0.83 using MRI slices and clinical data                                                                                                                                            \\
\hline
\end{tabular}
\end{sidewaystable}

Regarding the classification into one of the four RANO classes, there is still the need for an automated classification, which has been overlooked, probably due to the lack of publicly available data containing this classification. However, in late 2022, a new dataset (LUMIERE \cite{suter_lumiere_2022}) was published providing several advantages over other already available public datasets: i) it contains longitudinal patient data, ii) it provides several patient's clinical data including the RANO criteria and iii) it provides information on the acquisition parameters of each scan (e.g. MRI sequence, MRI scanner field strength, voxel size, etc). Although the dataset includes a smaller number of data points compared to the more recent editions of the BraTS dataset (usually over 1000), including a total of 638 timepoints, it provides a large amount of information for all data points, which is desirable, yet still uncommon.

\section{Methods}

\begin{figure}[h]
    \centering
    \includegraphics[width=0.9\textwidth]{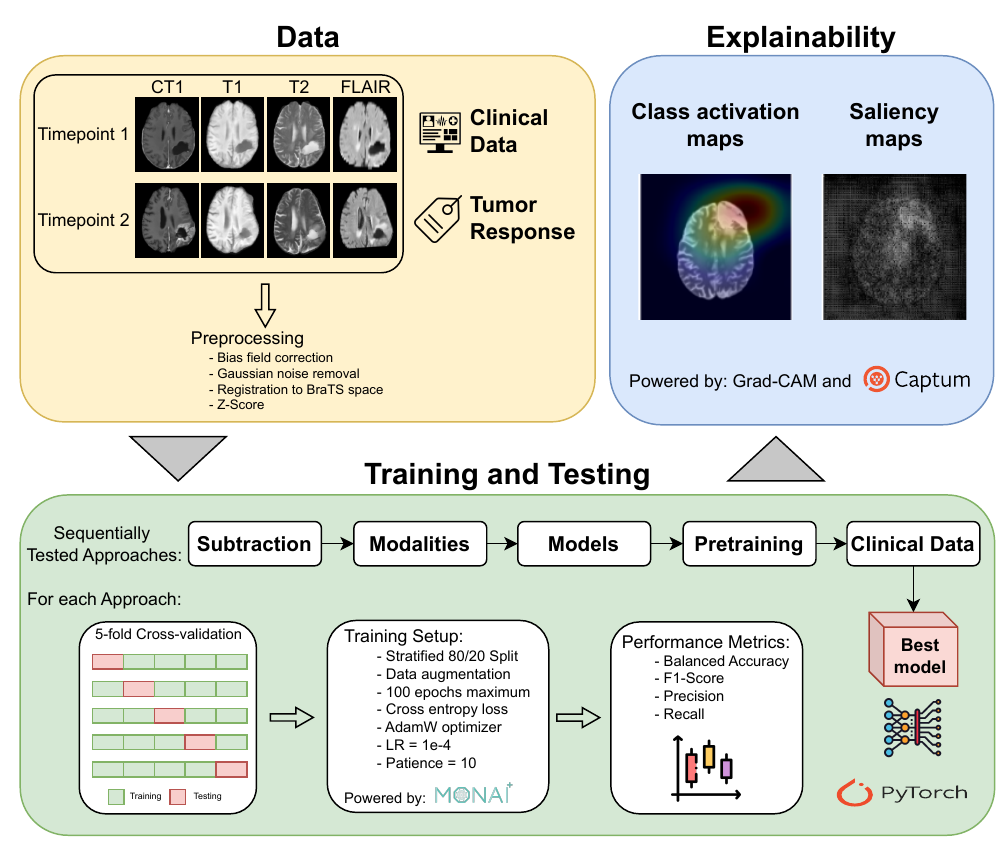}
    \caption{Overview of the pipeline of this project, including data preprocessing, the training setup, the sequential assessment of the different approaches, and finally, the use of different explainability methods on the model identified as best performant.}
    \label{pipeline_overview}
\end{figure}

Starting with a general overview of this work, the MRI and clinical data were first extracted from the LUMIERE dataset repository and preprocessed. After this, several methodological approaches were tested when implementing a DL pipeline with a common underlying training setup. The approaches were tested sequentially in a greedy method, choosing the best-performing option for each approach and keeping that option for the downstream tested approaches. After all approaches had been tested, the cumulative best-performing pipeline was submitted to two different explainability methods. An overview of the pipeline of this study is presented in Figure \ref{pipeline_overview}, and each step will be further detailed in the next subsections.

\subsection{Data}

\begin{figure}[h]
    \centering
    \includegraphics[width=0.95\textwidth]{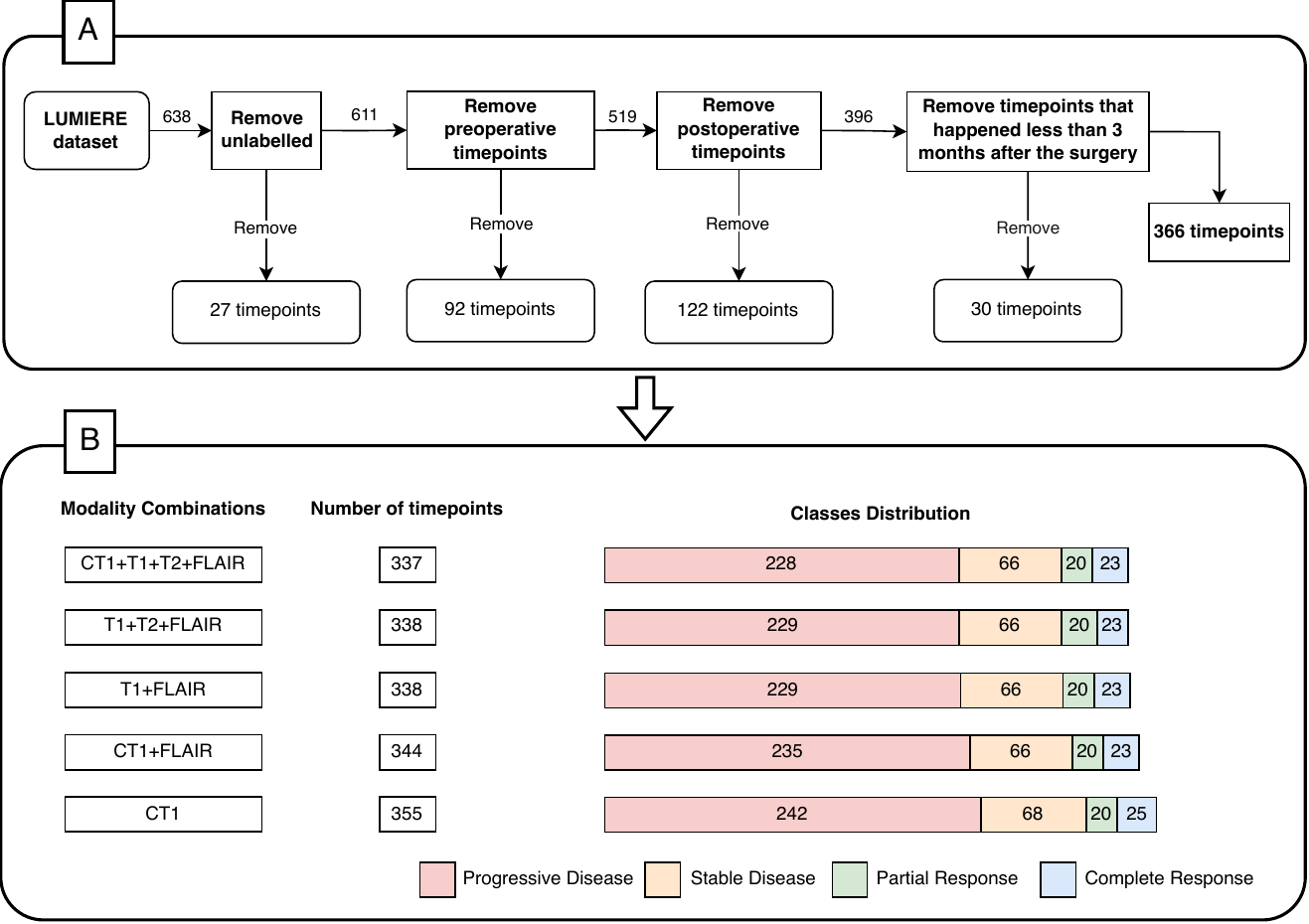}
    \caption{Diagram showing the two main parts of timepoint selection. (A) Firstly, the timepoints that were unlabeled or labeled as pre- or post-operative were removed, as well as any timepoint within 3 months of the surgery. In the end, 366 timepoints fit these restrictions, and include at least one kind of MRI modality. (B) Afterwards, since two consecutive timepoints are needed for the classification, only the timepoints that contained (at least) the desired combination of modalities in both timepoints were selected. As such, the different dataset sizes depending on the different combinations of modalities and the respective distribution of RANO classes are shown.}
    \label{filtragem_pacientes}
\end{figure}

The LUMIERE dataset \cite{suter_lumiere_2022} (acquired with the approval of the local ethics committee) is an open-source longitudinal dataset of glioblastoma (according to WHO 2016 guidelines) patients, comprising 91 patients with a total of 638 timepoints of MRI scanning. This dataset has multiparametric MRI data (T1-weighted contrast-enhanced - CT1, T1-weighted - T1W, T2-weighted - T2W, and T2 FLAIR), clinical data (sex, age at time of surgery, IDH mutation status, MGMT promoter status, and survival time), and the RANO criteria of each patient’s timepoint, when available. It is important to note that not all timepoints include all four imaging modalities. The demographic information of this dataset is presented in Table \ref{demographic}. Regarding clinical information, MGMT promoter methylation status and IDH mutation status were also provided, and their distribution is shown in Table \ref{demographic}. Concerning each timepoint’s classification, 253 were classified as PD, 97 as SD, 20 as PR, and 27 as CR, while the rest were either Pre-operative, Post-operative, or had no classification. As such, not all timepoints were used, since this work focused on the classification of RANO criteria. The pre- and post-operative timepoints were removed, as well as the unlabeled ones. Similarly, the ones that, despite having a RANO label assigned to them, happened less than 3 months after the surgery had taken place, were also removed as this is not in agreement with the recommended RANO guidelines \cite{wen_rano_2023} (see Figure \ref{filtragem_pacientes}A). Then, since two consecutive timepoints are required to classify the RANO label, it was necessary to select the ones that contain (at least) the same set of desired modalities. This results that, depending on the tested combination of modalities, the size of the usable dataset could be different (Figure \ref{filtragem_pacientes}B).

\begin{table}[h]
\caption{Demographic and clinical data on the LUMIERE dataset. PD=Progressive Disease, SD=Stable Disease, PR=Progressive Response, CR=Complete Response, NA=Not available}
\label{demographic}
\centering
\renewcommand*{\arraystretch}{1.3}
\begin{tabular}{|l|c|}
\hline
\multicolumn{1}{|c|}{Metric} & Value                                                \\ \hline
Age at surgery (years)       & 62 ± 10                                              \\ \hline
Sex                          & 51.6\% male (47M/44F)                                \\ \hline
Survival time (weeks)        & 83 ± 47                                              \\ \hline
IDH status                   & 57 wildtype | 10 IDH1 negative | 1 mutant | 23 NA    \\ \hline
MGMT gene promoter status    & 37 methylated | 43 not methylated | 11 NA            \\ \hline
RANO criteria classification & 253 PD | 97 SD | 20 PR | 27 CR                       \\ \hline
\end{tabular}
\end{table}

Before the images were inputted into the model, they were preprocessed: reorientation to Right-Anterior-Superior orientation (with nipype \cite{gorgolewski_nipype_2011}), N4 bias field correction (with ANTsPy \cite{tustison_antsx_2021}), Gaussian noise removal (with ANTsPy), registration to the SRI24 space (with DIPY \cite{garyfallidis_dipy_2014}), and Z-score intensity normalization. All details on the image pre-processing are available within the preprocessing script included in the code repository for this work, made openly available at https://github.com/anamatoso/RANO-classification. 

\subsection{Model Training}

The models were trained using the Pytorch Python package in conjunction with the MONAI Python package \cite{cardoso_monai_2022},  using the code provided in the repository mentioned above. Five-fold cross-validation was performed using an 80/20 stratified split, given that the classes were highly unbalanced (about 67\%, 20\%, 6\%, and 7\% for PD, SD, PR, and CR, respectively - Table \ref{demographic}). Additionally, due to the class imbalance, a weighted sampler was implemented at the data loader level so that each sample was assigned a weight inversely proportional to the number of instances in the ground truth class, as detailed below in Equation 1:

\begin{equation}
    W(s)=1-P(s)
\end{equation}

where W is the weight, s is the data sample, and P is the prevalence of the ground truth class of the sample s.
In each fold, the weights are randomly initiated using a Xavier normal function \cite{Glorot2010Mar}. Regarding the data transformations applied before training, the following data augmentation methods were used: random flips in all axes (probability of 50\%), random scaling of intensity (probability of 90\% by up to a factor of 0.1), random contrast adjustment (which alters the intensity of the image based on the gamma transform with a probability of 90\%), and random addition (probability of 90\%) of Gaussian noise (which adds Gaussian noise with 0 mean and 0.1 as standard deviation). All these augmentation methods are provided by the MONAI package.

Since preliminary tests showed that training stabilized very fast, as determined by the number of epochs required for the loss function to stabilize, early stopping was employed with 10 epochs as the patience value. The training process was allowed to last for a maximum of 100 epochs. Additionally, if the training loss did not diminish in a certain epoch, the learning rate was decreased by a factor of 10. The best model was selected according to the value of the test loss (the lower, the better).

In this study, we used the cross-entropy loss, applying a weighting per class equal to the inverse of the prevalence of that class in the training set, the Adam optimizer with a weight decay factor of 0.01, coupled with a learning rate of 1E-4, and a batch size of 4.

To assess the performance in each approach, we relied on the Balanced Accuracy, the Recall, the Precision, and the F1-Score, which are present in Equations 2 through 5, respectively:

\begin{equation}
    \textnormal{Balanced Accuracy}= \frac{1}{N}\sum_{i=1}^N \frac{{TP}_i}{{TP}_i+{FN}_i}
\end{equation}
\begin{equation}
    \textnormal{Recall}= \sum_{i=1}^N \frac{{TP}_i}{{TP}_i+{FN}_i}\frac{1}{n_i}
\end{equation}
\begin{equation}
    \textnormal{Precision}= \sum_{i=1}^N \frac{{TP}_i}{{TP}_i+{FP}_i}\frac{1}{n_i}
\end{equation}
\begin{equation}
    \textnormal{F1-score}= 2\times \frac{\textnormal{Precision} \times \textnormal{Recall}}{\textnormal{Precision} + \textnormal{Recall}}
\end{equation}

where TP=True Positives, FN=False Negatives, FP=False Positives, N=number of classes and n=number of samples

\subsection{Tested Approaches}

A baseline model was first defined, consisting of a Densenet architecture (densenet121 from MONAI) that takes 4 imaging modalities as input (CT1, T1W, T2W, and FLAIR) from 2 consecutive timepoints (therefore 8 images in total per data point). For each approach, the option that achieved the best performance was applied downstream in the subsequently tested approaches.

Firstly, we tested whether it would be beneficial to decrease the dimensionality of the input by two-fold. This meant providing as input only 4 subtraction images instead of the initial 8 MRIs, where each of those 4 images corresponded to the subtraction between consecutive timepoints for each imaging modality (labeled as Subtraction below).

Following this initial evaluation, different combinations of MRI modalities were tested to assess the impact of Modality selection, namely, the ones presented in Figure \ref{filtragem_pacientes}B. For each combination, data selection was made based on the timepoints of each patient, since the previous timepoint would also need to contain the same MRI modalities from that combination. So, for instance, for a model using only CT1 and FLAIR images as input, we selected only the timepoints of patients that had (at least) those two modalities in both the timepoint where the response to treatment was being assessed and the previous timepoint. Therefore, the size of the dataset available to train/test differed depending on which modality combination was being tested (decreasing from the total size found in Figure \ref{filtragem_pacientes}A). The sizes of the resulting datasets for all considered combinations of modalities is presented in Figure \ref{filtragem_pacientes}B. The reason for evaluating these specific combinations and not all possible options comes from different factors: to test the use of all the information available (CT1+T1W+T2W+FLAIR), to use only images that do not require contrast to accommodate patients who cannot undergo contrast injections (T1W+T2W+FLAIR), to use only the modality with the most pivotal role in treatment response assessment (CT1) \cite{wen_updated_2010,wen_response_2017}, to use only the two modalities that have been shown to be the most informative when segmenting tumors \cite{passarinho_automated_2023} (CT1+FLAIR), and the analogous of this last combination without the need for contrast injection (T1W+FLAIR).

Subsequently, five different DL model architectures were tested: three types of Densenets (Densenet121, Densenet169, and Densenet264) \cite{huang_radiological_2021}, a Visual Transformer (ViT) \cite{dosovitskiy_image_2021}, and a 3D adaptation of AlexNet \cite{krizhevsky_imagenet_2012} (AlexNet3D). These models all take into consideration the images as 3D volumes (thus having, for instance, 3D convolutions) instead of a single or a collection of 2D slices which would ignore the spatial dependencies between slices.

Then, the inclusion of three distinct pretraining tasks to the pipeline were tested. In this approach the initial weights of the model (whose architecture was defined by the previous approach) were determined by the result of each pretraining task The pretraining tasks were the following: 1) a self-supervised method applied to classify which rotation was applied to the image using the images from the LUMIERE dataset; 2) the use of the images from the MedMNIST dataset \cite{yang_medmnist_2023} to classify the organ present in the image; 3) the use of the checkpoint weights from MedicalNet \cite{chen_med3d_2019}.

Finally, the inclusion of clinical data (see Table \ref{demographic}) as an extra input to the model was also tested.

To evaluate between options considered for each methodological approach (Subtraction, Modality, Architecture, Pretraining, and Clinical Data), statistical tests were performed, using the Mann-Whitney U test when comparing only two options, the Kruskal-Wallis test when comparing between multiple groups (each group corresponding to a different approach), and the post-hoc Dunn test when appropriate. For all statistical tests a p-value of 0.05 was considered as significant.

\subsection{Explainability}
DL is often considered to operate as a black box. However, especially in the context of clinical applications, it is crucial to learn the reasoning of a model when it makes a particular prediction. As such, two explainability methods were applied to analyze the model that performed best after deciding on all the tested approaches.

The first explainability method employed was Grad-CAM \cite{selvaraju_grad-cam_2020} (using the Python package https://github.com/jacobgil/pytorch-grad-cam), which is a subtype of class activation maps (CAM). Grad-CAM calculates the weighted average of the feature maps of a certain model layer (usually the last convolutional layer) where the weights are the gradients of the model with respect to that layer, thus Grad-CAM. The Grad-CAM heatmaps yield a coarse localization of the regions that the model finds important when making a prediction.

The second explainability method used was Saliency maps using the Captum Python package \cite{kokhlikyan_captum_2020}. These maps are determined by calculating the voxel-wise gradients with respect to the input, which results in a much more granular effect of each voxel on the output of the model.

\section{Results}

\subsection{Tested Approaches}

\begin{figure}[ht]
    \centering
    \includegraphics[width=0.95\linewidth]{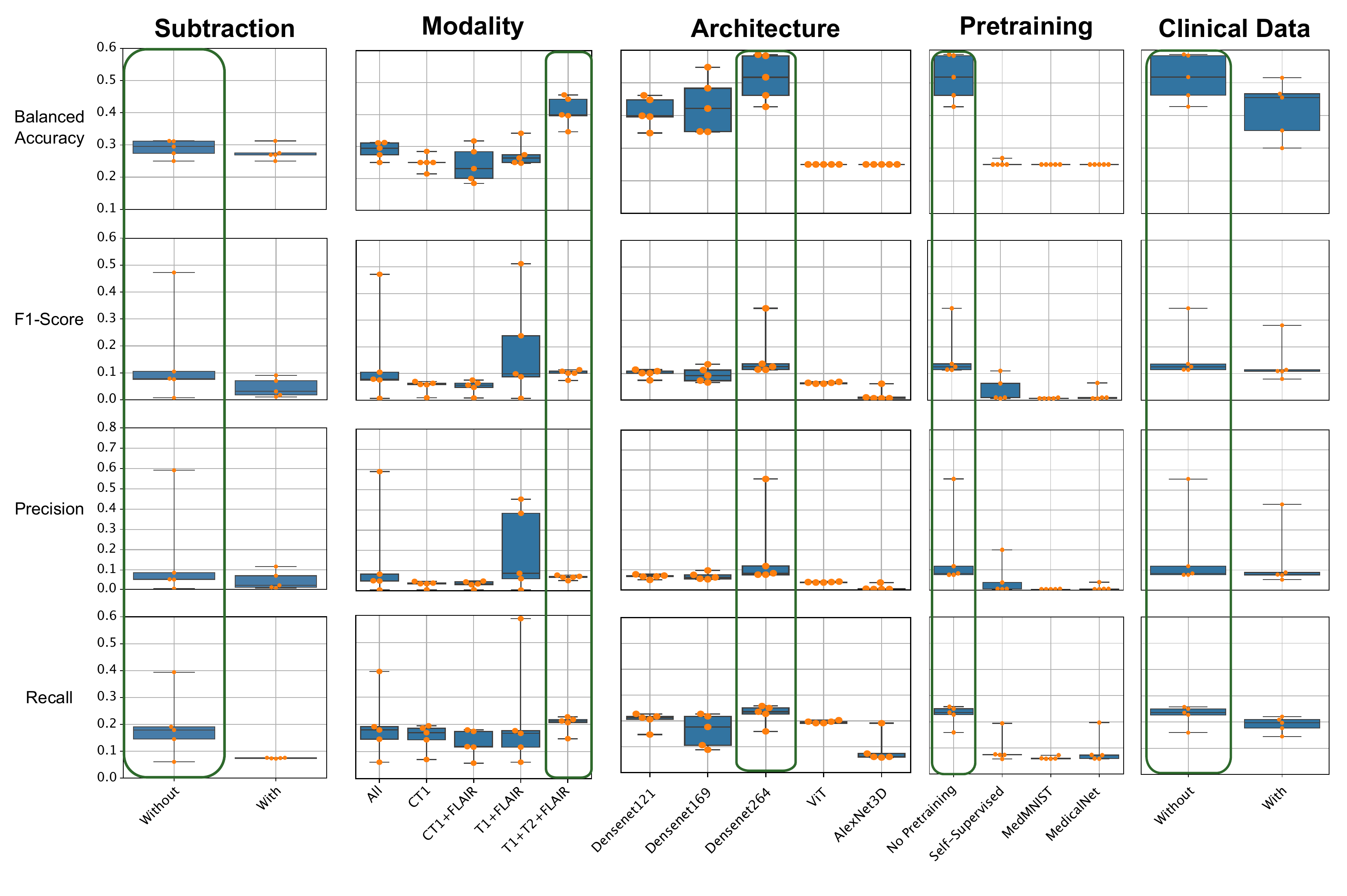}
    \caption{Impact of the different approaches on the models’ performance metrics (Balanced Accuracy, F1-Score, Precision, and Recall). The box plots represent the distributions of the performance metrics across folds. The green rectangles identify the most successful option for each approach. This option was applied to the pipeline when testing the subsequent approaches on the right. ViT=Visual Transformer.}
    \label{boxplots}
\end{figure}

In Figure \ref{boxplots}, the results for each evaluation metric are shown for all sequentially tested approaches, and Table \ref{pvalues} displays the results for the respective statistical tests performed.

\begin{longtable}{|c|c|c|c|c|c|c|}
\caption{P-values for the statistical tests performed comparing the different options for each approach (Mann-Whitney U for the Subtraction and the Clinical Data strategy, and Kruskal-Wallis test for the others), including post-hoc tests (Dunn test) when appropriate. Values are presented in bold if $p<0.05$, in which case the differences were considered significant. ViT=Visual Transformer.\label{pvalues}}\\ 
\hline
Strategy                     & Test                                                                            & Post-hoc Tests Pairs             & \begin{tabular}[c]{@{}c@{}}Balanced \\Accuracy\end{tabular} & F1-Score        & Precision       & Recall           \endfirsthead 
\hhline{|=======|}
Subtraction                  & Mann-Whitney U                                                                  & -                                & 0.53                                                        & 0.31            & 0.69            & 0.14             \\* 
\hhline{|=======|}
\multirow{11}{*}{Modalities} & Kruskal-Wallis                                                                  & -                                & \textbf{0.0074}                                             & \textbf{0.049}  & 0.045           & 0.21             \\* 
\cline{2-7}
                             & \multirow{10}{*}{\begin{tabular}[c]{@{}c@{}}Post-hoc \\Dunn Tests\end{tabular}} & All x CT1                        & 1                                                           & 1               & 1               & 1                \\* 
\cline{3-7}
                             &                                                                                 & All x CT1+FLAIR                  & 1                                                           & 1               & 0.94            & 1                \\* 
\cline{3-7}
                             &                                                                                 & All x T1+FLAIR                   & 1                                                           & 1               & 1               & 1                \\* 
\cline{3-7}
                             &                                                                                 & All x T1+T2+FLAIR                & 0.58                                                        & 1               & 1               & 1                \\* 
\cline{3-7}
                             &                                                                                 & CT1 x CT1+FLAIR                  & 1                                                           & 1               & 1               & 1                \\* 
\cline{3-7}
                             &                                                                                 & CT1 x T1+FLAIR                   & 1                                                           & 0.65            & 0.43            & 1                \\* 
\cline{3-7}
                             &                                                                                 & CT1 x T1+T2+FLAIR                & \textbf{0.016}                                              & 0.23            & 0.32            & 1                \\* 
\cline{3-7}
                             &                                                                                 & CT1+FLAIR x T1+FLAIR             & 1                                                           & 0.53            & 0.32            & 1                \\* 
\cline{3-7}
                             &                                                                                 & CT1+FLAIR x T1+T2+FLAIR          & \textbf{0.012}                                              & 0.18            & 0.23            & 0.20             \\* 
\cline{3-7}
                             &                                                                                 & T1+FLAIR x T1+T2+FLAIR           & 0.12                                                        & 1               & 1               & 1                \\* 
\hhline{|=======|}
\multirow{11}{*}{Models}     & Kruskal-Wallis                                                                  & -                                & \textbf{2.6e-4}                                             & \textbf{5.9e-4} & \textbf{4.7e-4} & \textbf{0.011}   \\* 
\cline{2-7}
                             & \multirow{10}{*}{\begin{tabular}[c]{@{}c@{}}Post-hoc \\Dunn Tests\end{tabular}} & Densenet121 x Densenet169        & 1                                                           & 1               & 1               & 1                \\* 
\cline{3-7}
                             &                                                                                 & Densenet121 x Densenet264        & 1                                                           & 1               & 1               & 1                \\* 
\cline{3-7}
                             &                                                                                 & Densenet121 x ViT                & 0.35                                                        & 0.64            & 0.77            & 1                \\* 
\cline{3-7}
                             &                                                                                 & Densenet121 x AlexNet3D          & 0.35                                                        & \textbf{0.040}  & 0.60            & 0.067            \\* 
\cline{3-7}
                             &                                                                                 & Densenet169 x Densenet264        & 1                                                           & 1               & 1               & 1                \\* 
\cline{3-7}
                             &                                                                                 & Densenet169 x ViT                & 0.14                                                        & 0.93            & 0.64            & 1                \\* 
\cline{3-7}
                             &                                                                                 & Densenet169 x AlexNet3D          & 0.14                                                        & 0.067           & \textbf{0.045}  & 0.74             \\* 
\cline{3-7}
                             &                                                                                 & Densenet264 x ViT                & \textbf{0.0018}                                             & \textbf{0.040}  & 0.30            & 1                \\* 
\cline{3-7}
                             &                                                                                 & Densenet264 x AlexNet            & \textbf{0.0018}                                             & \textbf{9.1e-4} & \textbf{7.6e-4} & \textbf{0.0072}  \\* 
\cline{3-7}
                             &                                                                                 & ViT x AlexNet                    & 1                                                           & 1               & 1               & 0.70             \\* 
\hhline{|=======|}
\multirow{7}{*}{Pretraining} & Kruskal-Wallis                                                                  & -                                & \textbf{9.5e-4}                                             & \textbf{0.0091} & \textbf{0.018}  & \textbf{0.014}   \\* 
\cline{2-7}
                             & \multirow{6}{*}{\begin{tabular}[c]{@{}c@{}}Post-hoc \\Dunn Tests\end{tabular}}  & No Pretraining x Self-Supervised & \textbf{0.031}                                              & \textbf{0.017}  & \textbf{0.032}  & \textbf{0.025}   \\* 
\cline{3-7}
                             &                                                                                 & No Pretraining x MedMNIST        & \textbf{0.0056}                                             & 0.11            & 0.26            & 0.075            \\* 
\cline{3-7}
                             &                                                                                 & No Pretraining x MedicalNet      & \textbf{0.0015}                                             & \textbf{0.016}  & \textbf{0.027}  & \textbf{0.033}   \\* 
\cline{3-7}
                             &                                                                                 & Self-Supervised x MedMNIST       & 1                                                           & 1               & 1               & 1                \\* 
\cline{3-7}
                             &                                                                                 & Self-Supervised x MedicalNet     & 0.91                                                        & 1               & 1               & 1                \\* 
\cline{3-7}
                             &                                                                                 & MedMNIST x MedicalNet            & 1                                                           & 1               & 1               & 1                \\ 
\hhline{|=======|}
Clinical Data                & Mann-Whitney U                                                                  & -                                & 0.056                                                       & \textbf{0.0079} & \textbf{0.0079} & 0.60             \\
\hline
\end{longtable}

Firstly, the impact of subtracting the images of each modality was tested as a way of reducing the input size. As can be seen in Figure \ref{boxplots}, there appears to be a performance improvement when no subtraction was applied. Nevertheless, this performance improvement was not statistically significant as evaluated using the Mann-Whitney U test presented in Table \ref{pvalues}. Therefore, henceforth, no subtraction was applied to the input images.

Afterward, regarding the different combinations of modalities tested, two combinations seem to outperform the others: the one that uses T1W+FLAIR as input, which resulted in the largest overall Precision and F1-Score, and the model that uses T1W+T2W+FLAIR as input, whose values of Balanced Accuracy and Recall were higher than the rest. A Kruskal-Wallis test was performed to analyze the statistical differences between inputting distinct modality combinations. These results can be seen in Table \ref{pvalues}. In this statistical test, p-values of 0.0074, 0.049, and 0.045 were found for Balanced Accuracy, F1-Score, and Precision, respectively, which prompted a post-hoc analysis that revealed a significant difference in Balanced Accuracy between the T1W+T2W+FLAIR combination and the CT1 (p=0.016), and between the T1W+T2W+FLAIR combination and the CT1+FLAIR combination (p=0.012). Given these results and the fact that T1W+FLAIR is a subset of T1W+T2W+FLAIR, the model that takes as input T1W+T2W+FLAIR was chosen for subsequent evaluations. 

Concerning the different model architectures tested, the Kruskal-Wallis test detected a significant difference between the groups in all metrics (p<0.001 for Balanced Accuracy, F1-Score, and Precision, and p=0.012 for Recall). Overall, when looking at the metrics’ box plots (especially the Balanced Accuracy one), all Densenets showed better performance than the ViT and the AlexNet3D. The results also showed improved performance as the network’s complexity increased. In fact, the Densenet264 reached a median Balanced Accuracy of 51\% (achieving over 55\% in two of the folds) and outperformed the other models in the remaining metrics. Statistically speaking, Densenet264 shows significant differences in all metrics when compared to the AlexNet3D (p<0.01 for all metrics) and in all except Recall when compared to the ViT architecture (p<0.01 for Balanced Accuracy and p=0.023 for both F1-Score and Precision). Hence, the Densenet264 was the model architecture chosen to move forward.

As for the different pretraining approaches tested, the Kruskal-Wallis test revealed a significant difference between the groups (p<0.01 for Balanced Accuracy, F1-Score, and Recall and p<0.02 for Precision). When performing the post-hoc tests, the option where no pretraining was applied was significantly better than the Self-supervised option in all metrics (p<0.05) and it also had a significantly higher Balanced Accuracy when compared to the MedMNIST and MedicalNet options (p<0.02). Given these results, the best option in this case was not to perform any pretraining beforehand.

In what concerns the use of clinical data as input, the Balanced Accuracy seemed to be higher when clinical data was not fed into the model, whereas the other metrics appeared to have little to no change, yet no significant results were found. As such, we decided to continue with the model that did not use clinical data as input.

Finally, to sum up, the cumulative options for each approach that yielded the model that performed best were as follows: no subtraction applied to the input images, the use of T1-weighted, T2-weighted, and FLAIR images, selection of the Densenet264 model architecture from MONAI, no pretraining of the model and no clinical data used as input.

\subsection{Explainability}

\begin{figure}[h]
    \centering
    \includegraphics[width=0.9\textwidth]{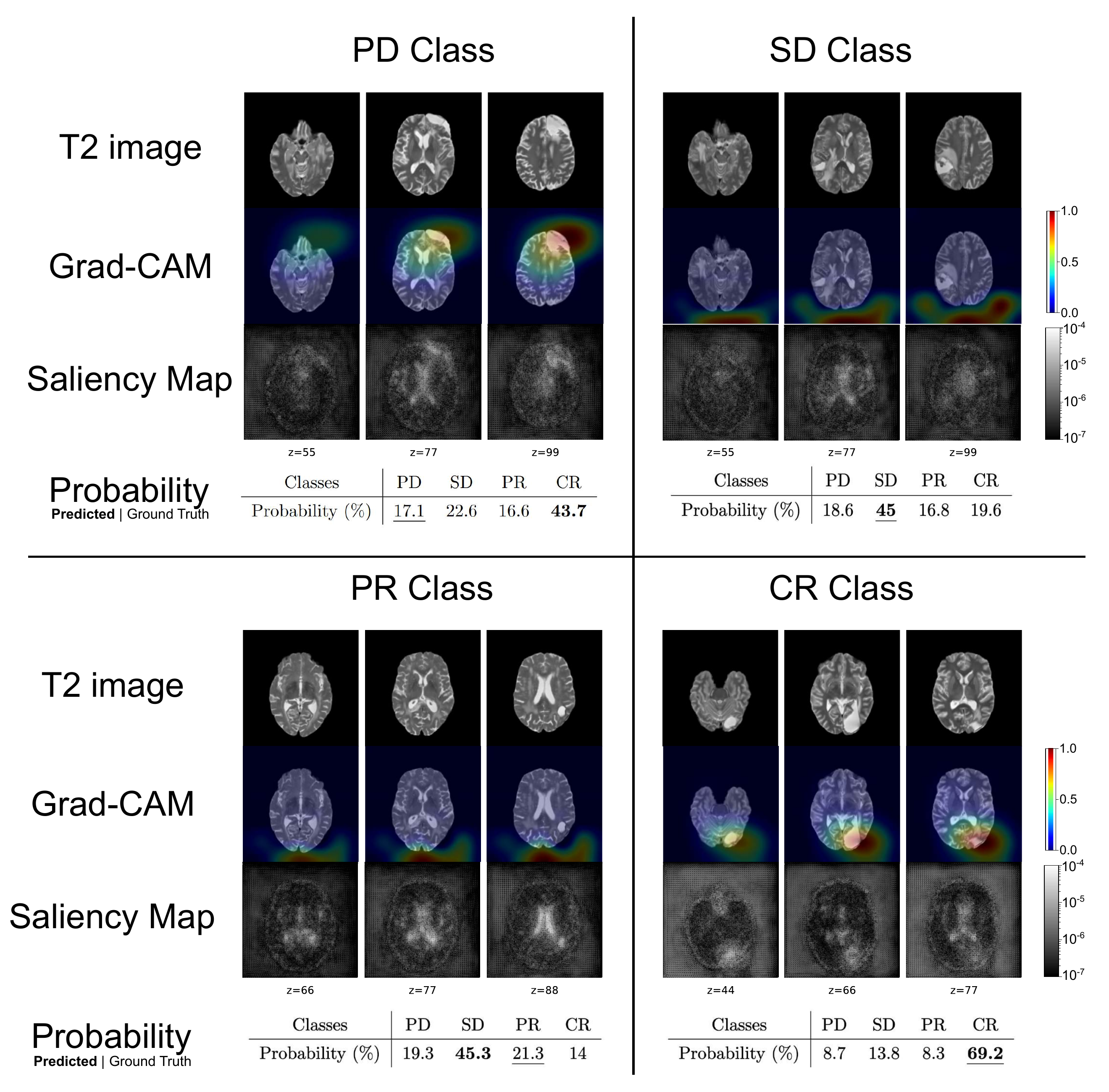}
    \caption{Reference T2-weighted images, Grad-CAM overlayed on top of T2-weighted image, and Saliency Maps for three slices of one example from each of the four RANO classes, with the corresponding color bars. The best-performing model was used. The probability of prediction for each class is also presented for each specific example. PD=Progressive Disease; SD=Stable Disease; PR=Progressive Response; CR=Complete Response.}
    \label{explainability}
\end{figure}

In this section, the results from both explainability methods described above will be explored for the best cumulative model that was defined in the previous section. In Figure \ref{explainability}, four examples of the explainability heatmaps can be seen, as well as the class prediction probability for each example. The results for the remaining data of the test set can be seen in the Supplementary Material.

Saliency maps seem to more easily show the tumor region, unlike Grad-CAM, which only highlights the tumor region in the PD and CR examples for the ground truth class. In fact, for the SD and PR examples, Grad-CAM does not even highlight any brain region. Nevertheless, in the PD class (ground truth), the class activates some non-tumorous brain regions.

Regarding the prediction probabilities, only in the SD and CR classes does the higher probability correspond to the ground truth class. The CR class has the higher prediction probability, achieving almost 70\%. Moreover, it was observed that most tumors in the CR class were located in the occipital lobe. The prediction probabilities of the whole test set can be seen in Table S1 of the Supplementary Material.

\section{Discussion}

This work systematically analyzed different DL approaches for the RANO classification of glioblastomas’ response to treatment. The results showed there was a higher performance when T1-weighted, T2-weighted, and FLAIR images were inputted into a Densenet264 architecture. In addition, two explainability methods were explored to see which regions of the images were considered most important for the classification by the model.

\subsection{Tested Approaches}

Previous studies have used subtraction of MR images acquired at different timepoints to enhance lesion detection, for instance in multiple sclerosis and breast cancer, with promising results showing improved detection of lesions\cite{moraal_subtraction_2009, loizidou_automated_2020}. Nevertheless, in our study, we found that performing the subtraction of MR images did not improve the performance of the model, suggesting that there may be image features in the raw images that are useful for the classification that are eliminated in the subtracted images. In addition, subtraction images may have errors coming from the registration process and the subtraction itself, which might add noise to the data.

Concerning the different MRI modality combinations tested, it is interesting to note that using all four modalities available did not provide the highest performance. Instead, the highest performance was achieved when all modalities were used but the CT1. While this is the modality that yields a better visualization of the tumor core, these results suggest that it may not be as important for RANO classification. We hypothesize that the fact that CT1 enhancement regions are typically smaller than T2W/FLAIR enhancement regions may contribute to it being disregarded as important when compared to the T2W/FLAIR modality. Additionally, adding the T2W image to the T1W+FLAIR input combination (leading to T1W+T2W+FLAIR) seems to improve Balanced Accuracy whilst diminishing Precision and F1-Score, which might imply that this modality is important for the distinction of RANO labels, especially progression versus non-progression (since the higher Precision and F1-Score are due to an overclassification of PD class). T2W images have proven to be important when classifying MGMT promoter methylation \cite{huang_radiological_2021, villanueva-meyer_artificial_2024} and the codeletion of chromosome arms 1p and 19q \cite{yogananda_novel_2020, han_non-invasive_2018}. Both biomarkers play a part in treatment response in gliomas. Nevertheless, more advanced modalities like diffusion and perfusion imaging have been shown to be useful for detecting pseudoprogression and for prognostication \cite{villanueva-meyer_artificial_2024, kim_incorporating_2019}. In addition, there has been an emergence of metabolic imaging, such as positron emission tomography (PET), to study gliomas for both diagnosis and treatment assessment \cite{hoggarth_clinical_2024}, in some cases revealing to be up to par with MRI-based methods \cite{lohmann_predicting_2018}. Although the dataset used in this work does not include these modalities, preventing the validation of models that integrate them, these different types of medical images should, if possible, be included in any project aiming for the longitudinal assessment of gliomas.

Moreover, the Densenet was the architecture that performed best. Densenets connect each layer to every other layer, thus reducing the vanishing gradient issue common to standard convolutional neural networks, with a relatively low number of parameters \cite{huang_radiological_2021}. These networks have shown good performance in a variety of tasks from the classification of IDH mutation \cite{liang_multimodal_2018} to MGMT promoter methylation status \cite{yogananda_mri-based_2021} with varied degrees of success. Other networks, such as Visual Transformers and AlexNet, have presented high performances for the classification of 2D images, where pre-trained models can be leveraged \cite{yang_glioma_2018, mzoughi_vision_2024}. Furthermore, ensemble models - that combine the outputs of a set of models - seem to outperform “single-model” models at many tasks \cite{patil_ensemble_2023, kang_mri-based_2021, hossain_vision_2024} and would thus be a promising direction for any future work.

In addition, the performance decreased when a pretraining method was used. This suggests that the information learned on the pretraining tasks was not helpful for the classification of the RANO criteria, which in turn means that there might be more complex factors in play for the RANO classification that the pretraining tasks did not capture. However, this does not mean that other types of pretraining tasks could not eventually improve training.

Moreover, it could appear strange that, by inputting clinical data, the Balanced Accuracy diminishes. However, a sampling bias may be happening at the level of some of the genetic information that is present in the clinical data of this dataset, especially the IDH mutation and the MGMT promoter status. Due to the heterogeneity of the tumor, the site where the biopsy was done might not represent the genetic profile of that tumor \cite{booth_imaging_2022}, resulting in a sampling bias that can be introducing noisy information into the training data. This issue will, in turn, lead to wrong information being inputted into the model, which will hinder learning. Furthermore, this dataset has, for each patient, only one genetic test in which IDH mutation and MGMT mutation were evaluated. Since this is a longitudinal dataset, and given the dynamic nature of glioblastoma, the recorded genetic mutation may not be reliable, especially for timepoints distant from the first surgery date. In addition, it can be seen that the distribution of IDH mutation has many missing data points and is mostly IDH wildtype. On the other hand, the MGMT mutation status distribution is quite balanced, although it has a few missing data points as well. These distributions, the first one being quite non-informative, and the second one being rather dispersed, could explain the lower performance of the model when clinical data was included as they do not seem to reflect the label (RANO criteria) distribution.

\subsection{Explainability}

Regarding the explainability methods, they present a heatmap that can be interpreted as to which were the voxels of interest when making the prediction. In this work, the results proved to be rather volatile, highlighting the tumor region only in some cases, with no apparent correlation to whether the model’s prediction was correct or not. This happened especially in Grad-CAM since in the Saliency Maps, the tumor regions seem to be mostly identified. We hypothesize that this might be because Grad-CAM calculates the weighted average of the feature maps of the last convolutional layer and thus 1) inherits the resolution of that layer, which is much lower than the input image, 2) can also represent abstract features that do not depend directly on the image intensity and 3) by performing the average of the feature maps a dilution of the maps might be happening. 

To combat this issue, a segmentation process could be placed upstream in this pipeline to restrict the classification model only to the tumor region, as other researchers have done \cite{naser_brain_2020, vafaeikia_mri-based_2024}. Nevertheless, it has been shown that the overlap of the two highlighted regions of the two methods might be related to the correctness of the prediction, meaning that if the two methods highlight the same region, it could indicate that the model correctly predicts the class \cite{gulum_improved_2021}. However, though these methods can display which were the voxels of interest  when it makes a prediction, they cannot tell why  that region was highlighted, which in a clinical context is just as important \cite{lipkova_artificial_2022}. As such, the clinical implementation of these models must be made as a support tool for medical decision-making. Furthermore, and as it was observed, tumors in the occipital region have been linked to longer survival, possibly due to their relative ease of treatment \cite{nizamutdinov_prognostication_2018}.

\subsection{Limitations and Innovative Aspects}

This work had some limitations, namely the limited amount of data available for training, which was evidently not enough for the model to fully understand the characteristics of the tumor that led it to decide on a certain RANO classification. Well-annotated, large, and open datasets are still scarce for this application, often leading researchers to use smaller (mostly private) datasets \cite{moassefi_deep_2022, jang_prediction_2018,xue_prediction_2021, kanas_learning_2017}, which hinders the learning performance of DL models and diminishes the trained model’s robustness and generalizability. To tackle this issue, federated learning techniques could be used where, instead of centralizing the data (which has its privacy concerns), each medical institution would train a model locally and share only the weight updates of the model \cite{guan_federated_2024}. By proceeding in this way, the patient’s data would not leave each institution’s premises, and a more generalizable model could be determined. Concerning the acquisition protocol, the dataset has MR images acquired from different scanners, so data harmonization (for instance, ComBat \cite{fortin_harmonization_2018}) or at least intensity normalization (Z-score or histogram matching \cite{nyl_standardizing_1999}, for example) could potentially improve the results.

Additionally, given the complexity of this classification task and the difficulty medical professionals sometimes face in assigning rigid class labels, perhaps a soft labeling system (in which each data sample has a probability of belonging to each class), or the assignment of a confidence level to each data sample could be more appropriate, though this has its challenges. Moreover, independent validation of these results should be performed to verify the generalizability of the pipeline. Furthermore, as future work, a segmentation network, such as nnU-Nets \cite{Isensee2021Feb} or Swin-UNETR \cite{Tang2022}, could be used as a first step to restrict the analyzed volume to only the tumor region thus preventing the model from taking into consideration non-tumorous regions.

Concerning the innovative aspects of this work, it must be highlighted that training DL models to classify brain tumors into these four RANO criteria has, to the authors’ knowledge, never been done before, which is possibly because of the lack of open datasets that also make this categorization available. There is already a paucity of longitudinal tumor datasets \cite{suter_lumiere_2022, bakas_university_2022, scarpace_cancer_2016, mamonov_data_2016, national_cancer_institute_clinical_proteomic_tumor_analysis_consortium_cptac_clinical_2023}, and among those, we were only able to find one that included the RANO labels (the LUMIERE dataset). Nevertheless, despite the results being below clinical standards, hovering around 50\% (albeit above random since in a four-class problem, chance accuracy would be 25\%), this work breaks ground in classifying treatment response using RANO criteria using DL and thus benchmarks this task for future studies on the topic. Still, further work is needed in this field of research, potentially exploring other methodologies/frameworks, such as using radiomics and traditional machine learning, or by coupling large language models and clinical notes.

\section{Conclusion}

The main contribution of this paper is the systematic study of different approaches for RANO criteria classification using DL and glioblastoma MR images, which had never been done. Although the results were more modest than usually expected from DL models, better-than-chance results were obtained. By using a Densenet264 model with three different MRI modalities in two consecutive timepoints, we achieved a median Balanced Accuracy of 51\% (reaching over 55\% in two of the folds). This work thus shows how complex the classification into one of the four RANO criteria is, which probably comes from both the heterogeneity of the tumors themselves and the challenges of collecting sufficiently large datasets of longitudinal glioma MR data. Nevertheless, in this work, only images (and clinical information at times) were used for the classification. However, in a clinical setting, radiologists have more information at hand, namely the patient history, clinical reports, and lab data that could help improve the performance of more complex DL models to be developed in the future. In conclusion, although the proposed DL models are not yet fit for clinical use, this methodology shows potential and should continue to be explored in future work.

\section*{Ackowledgements}
This work was supported by the Portuguese Science Foundation through grants 2023.03810.BDANA, 2022.13185.BD, and UI/BD/154928/2023. This work was also supported by LARSyS FCT funding (DOI: 10.54499/LA/P/0083/2020, 10.54499/UIDP/50009/2020, and 10.54499/UIDB/50009/2020).

\bibliographystyle{unsrt}  
\bibliography{OPTICAmeetings}  
\end{document}